\documentclass[%
prl, twocolumn,
superscriptaddress,
 amsmath,amssymb,
 aps, 
]{revtex4-1}

\usepackage{times}
\usepackage{graphicx}
\usepackage{dcolumn}
\usepackage{bm}
\usepackage{chemarrow}

\usepackage{dcolumn}
\usepackage{amsthm}
\usepackage[titletoc]{appendix}
\usepackage{titlesec}
\usepackage{enumerate}

\makeatletter

\newcommand{\RNum}[1]{\uppercase\expandafter{\romannumeral #1\relax}}
\makeatother

\theoremstyle{definition}

\theoremstyle{plain}

\theoremstyle{remark}

\begin{document}

\preprint{APS/123-QED}

\title{Universal Relation Between Thermodynamic Driving Force and One-Way Fluxes in a Nonequilibrium Chemical Reaction with Complex Mechanism}

\author{Yongli Peng}
\email{yonglipeng@pku.edu.cn}
\affiliation{School of Mathematical Science,Peking University, China}
\author{Hong Qian}
\email{qian@amath.washington.edu}
\affiliation{%
Department of Applied Mathematics, University of Washington, Seattle, USA
}
\author{Daniel A. Beard}
\email{beardda@umich.edu}
\affiliation{
Department of Molecular and Integrative Physiology, University of Michigan, USA
}
\author{Hao Ge}%
 \email{haoge@pku.edu.cn}
\affiliation{%
Beijing International Center for Mathematical Research  (BICMR) and Biomedical Pioneering Innovation Center (BIOPIC), Peking University, China
}

\par
\begin{abstract}
In nonequilibrium chemical reaction systems, a fundamental relationship between unbalanced kinetic one-way fluxes and thermodynamic chemical driving forces is believed to exists. However this relation has been rigorously demonstrated only in a few cases in which one-way fluxes are well defined. In terms of its stochastic kinetic representation, we formulate the one-way fluxes for a general chemical reaction far from equilibrium, with arbitrary complex mechanisms, multiple intermediates, and internal kinetic cycles. For each kinetic cycle, the logarithm of the ratio of the steady-state forward and backward one-way fluxes is equal to the free energy difference between the reactants and products along the cycle. This fundamental relation is further  established for general chemical reaction networks with multiple input and output complexes. Our result not only provides an equivalent definition of free energy difference in nonequilibrium chemical reaction networks, it also unifies the stochastic and macroscopic nonequilibrium chemical thermodynamics in a very broad sense.
\end{abstract}

\pacs{Valid PACS appear here}
\maketitle


{\em Introduction.} In the terminology of classical 
mechanics, chemical kinetics and thermodynamics
correspond to {\em kinematics} and {\em dynamics},
respectively \cite{2016arXiv160508070Q,ComplexSci}. We show a universal relationship between kinetic
fluxes and thermodynamic driving forces in a nonequilibrium chemical reaction network (CRN), which are key concepts in the respective theories.  More specifically, we prove 
\begin{equation}\label{eqstar}
\Delta G = -k_BT\ln\left(\frac{J^+}{J^-}\right),    
\end{equation}
for a general, not necessarily elementary, chemical reaction, where $J^+$ and $J^-$ are forward and backward one-way fluxes and $\Delta G$ is the corresponding free energy difference.  Relation (\ref{eqstar}) has been established for several special situations in the past \cite{Hill1966Studies, Hill1980Free, Hill1975Stochastics, Briggs338}; the present work provides the most general theory to date.

Once relation (\ref{eqstar}) holds, the associated entropy production rate is immediately 
$$ \text{epr} = -J\Delta G = k_BT\big(J^+ - J^-\big)\ln\left(\frac{J^+}{J^-}\right) \geq 0,$$
in which $J = J^+-J^-$ is the net flux. The final equality in the above equation holds if and only if $J = \Delta G = 0$, which implies the principle of detailed balance in thermochemical equilibrium \cite{Lewis1925A}.  Eq. \ref{eqstar} is known to be closely related to the fluctuation theorem for entropy production \cite{CrookGE1999,PhysRevLett.74.2694,Qian2001Nonequilibrium,Hong2006Generalized}.

For any stochastic reversible elementary reaction, one-way fluxes in both directions are well defined; they equal to the forward and backward reaction rates. For a complex CRN composed of many reversible elementary reactions, however, so far one-way flux in a nonequilibrium steady state (NESS) has been clearly defined only when all reactions are first-order or pseudo-first-order, i.e. the kinetics is linear. How to generalize Eq. \ref{eqstar} to nonlinear, non-elementary reactions has been an open question \cite{Beard2007Relationship}.

The concept of one-way flux, introduced and extensively studied in \cite{Hill1980Free} for linear CRN, relies on the notion of mean first passage time in stochastic processes.  Hill's approach cannot be generalized to nonlinear chemical kinetics. This is the main obstacle for introducing one-way flux into a general stochastic CRN, which had cast doubt on the universality of Eq. \ref{eqstar} \cite{Ross2008}.

In this letter, we consider the general CRN in a NESS sustained by a chemostat \cite{Rao2016Nonequilibrium}. The one-way flux of each reaction cycle is defined through the cycle fluxes in the counting space of the corresponding chemical-master-equation model \cite{mcquarrie_1967} describing the stochastic kinetics of molecular numbers. Our theory is based on the general mathematical results on cycle fluxes of a Markov process \cite{Qian1981Markov,Jiang20042}: We prove Eq. \ref{eqstar} for each reaction cycle by summing over all the corresponding stochastic cycles in the counting space and derive the entropy production rate in terms of the one-way cycle fluxes.  Finally, we generalize all these results to the most general cases with the presence of multiple material reservoirs.


{\em A Chemical Reaction with Complex Mechanisms.} A general CRN in a
continuously stirred tank reactor consists of a set of species, $X_1, X_2, \ldots, X_{N_1}, Y_1, Y_2, \ldots, Y_{N_2}$, and a set of $M$ reactions between them,
$R_{1}, R_2, \ldots, R_{M}$. Species are further classified into \textit{internal} $Y$'s
and \textit{external} $X$'s. The concentrations of all the species $X$'s are clamped at constant levels.
All reactions are stochastic, elementary and reversible, under isothermal and isobaric conditions with fixed volume $V$. A closed CRN has $N_1=0$.

\begin{figure}[htb]
\[
 \includegraphics[width=.45\textwidth]{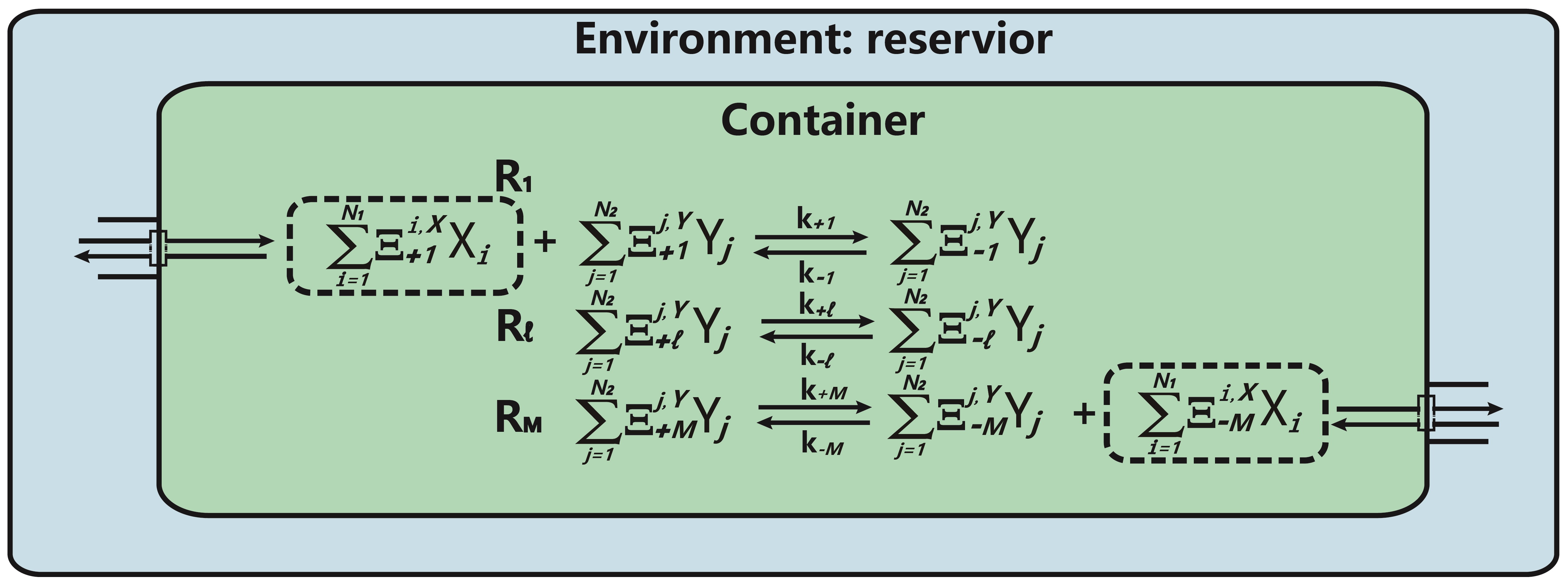}
\]
\caption{Representation of a general CRN. $X$'s are chemostated species and $Y$'s are internal species,  e.g., intermediates in the complete mechanism for transforming $X$'s on the left to those on the right.}
\label{fig:1}
\end{figure}

The $M$ reactions, including those between the 
internal species and chemostated ones can be classified into the three groups:
\begin{eqnarray*}
R_1: && \sum_{j=1}^{N_2} \Xi_{+1}^{j,Y} Y_{j} + \sum_{i=1}^{N_1} 
    \Xi_{+1}^{i,X} X_{i} \ \rightleftharpoonsfill{22pt}\ \sum_{j=1}^{N_2} \Xi_{-1}^{j,Y} Y_{j},
\\
R_{\ell}: && \sum_{j=1}^{N_2} \Xi_{+\ell}^{j,Y} Y_{j} \ \rightleftharpoonsfill{22pt}\ \sum_{j=1}^{N_2} \Xi_{-\ell}^{j,Y} Y_{j},
\\
R_M: && \sum_{j=1}^{N_2} \Xi_{+M}^{j,Y} Y_{j} \ \rightleftharpoonsfill{22pt}\ \sum_{i=1}^{N_1} \Xi_{-M}^{i,X} X_{i} + \sum_{j=1}^{N_2} \Xi_{-M}^{j,Y} Y_{j};
\end{eqnarray*}
where $\ell=2,3,\ldots,M-1$. Here only the left-hand-side of reaction $R_1$ and the right-hand-side of reaction $R_M$ exchange materials between $X$ and $Y$. Hence this reaction has a single input complex and a single output complex (FIG. \ref{fig:1}). All the $Y$'s species are intermediates in the mechanistic details of the overall reaction, transforming $ \sum_{i=1}^{N_1} \Xi_{+1}^{i,X} X_{i}$ to $\sum_{i=1}^{N_1} \Xi_{-M}^{i,X} X_{i}$.

Introducing the stoichiometric matrices $\bold{\Xi}^X$
and $\bold{\Xi}^Y$ for $X$'s and $Y$'s respectively: 
\[
    \Xi_{i\ell}^X=\left\{ \Xi_{-\ell}^{i,X}-\Xi_{+\ell}^{i,X}\right\}_{N_1\times M},
    \Xi_{j\ell}^Y=\left\{ \Xi_{-\ell}^{j,Y}-\Xi_{+\ell}^{j,Y}\right\}_{N_2\times M}.
\]
The complete stoichiometric matrix, then, is $\bold{\Xi} = \left[\begin{array}{c}\bold{\Xi}^X\\\bold{\Xi}^Y\end{array}\right]$. Note in this notion only reactions $1$ and $M$ possess nonzero coefficients in $\bold{\Xi}^X$,
for external species.

\begin{figure}
\[
 \includegraphics[width=.45\textwidth]{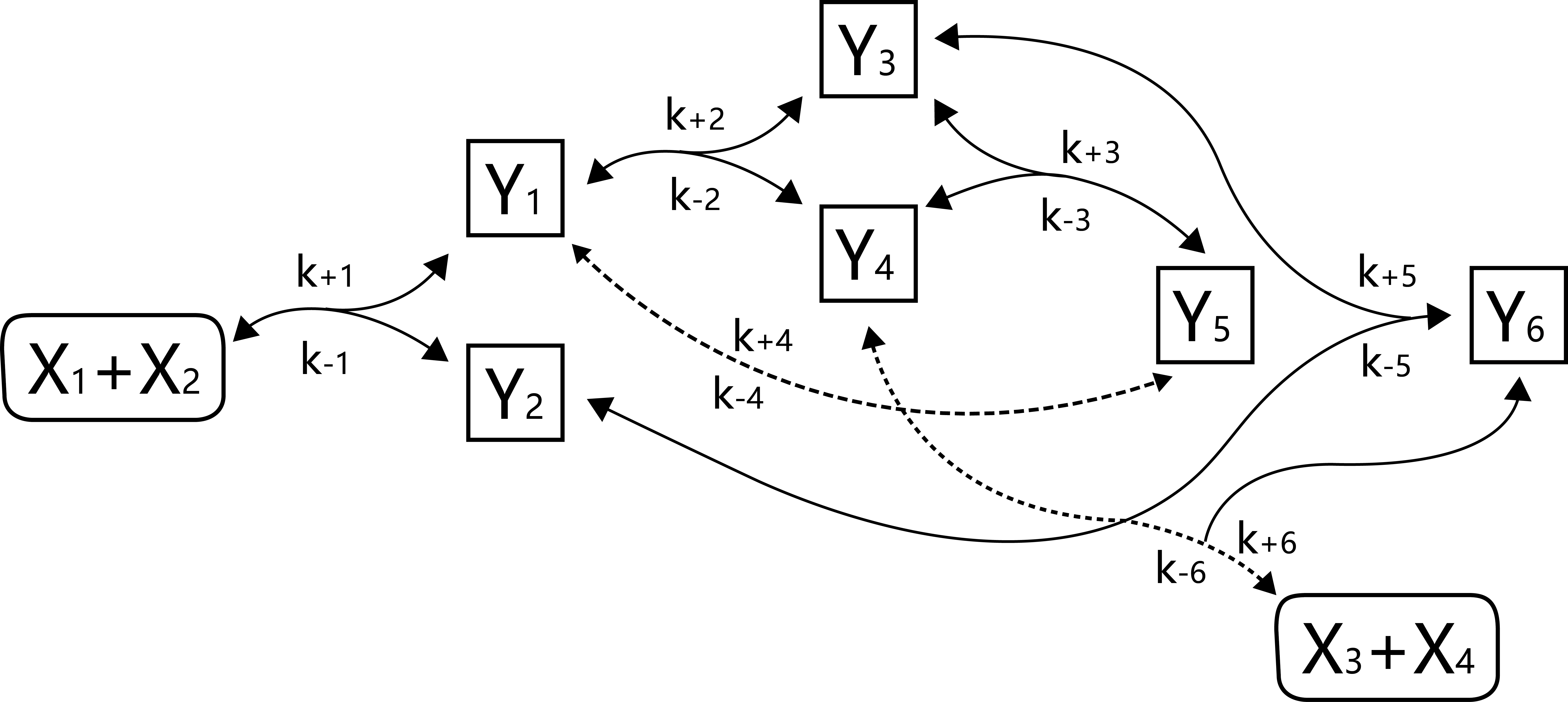}
\]
\caption{An example with single input complex and single output complex, in which the Wegscheider cycle condition requires that $k_2k_3k_{-4}=k_{-2}k_{-3}k_4$.}
\label{fig_example}
\end{figure}

FIG. $\ref{fig_example}$ is an example with the overall reaction $X_1 + X_2 \ \rightleftharpoonsfill{20pt}\ X_3 + X_4$. $X_1$, $X_2$, $X_3$, and $X_4$ are the external species,
with their concentrations kept as chemostated, and $Y_1$ to $Y_6$ are all
intermediates.

We shall use $x_i(t)$ and $y_j(t)$ to denote the concentrations of $X_i$ and $Y_j$, respectively, at time $t$.  Their macroscopic kinetics satisfies the rate equation:
\[
                  \frac{d y_j}{dt} = \sum_{\ell=1}^{M}
\Xi_{j\ell}^YJ^{\ell}, \quad   1\le j \le N_2,
\]
where $J^{\ell} = J^{+\ell} - J^{-\ell}$ is the net flux of the $\ell^{th}$ elementray reaction. 
$J^{\pm\ell}$ are the macroscopic one-way fluxes satisfying the law of mass action \cite{M.Feinberg,de2013non, pekavr2005thermodynamics}, i.e.
\[
            J^{\pm\ell} = k_{\pm\ell} \prod_{j=1}^{N_2}
{y_{j}}^{\Xi_{\pm\ell}^{j,Y}},
\]
except 
\[
           J^{+1} =  k_{+1} \prod_{i=1}^{N_1}
{x_{i}}^{\Xi_{+1}^{i,X}}\prod_{j=1}^{N_2}
{y_{j}}^{\Xi_{+1}^{j,Y}},
\]
\[
J^{-M} =  k_{-M} \prod_{i=1}^{N_1}
{x_{i}}^{\Xi_{-M}^{i,X}}
\prod_{j=1}^{N_2}
{y_{j}}^{\Xi_{-M}^{j,Y}}.
\]
$k_{\pm\ell}$ are called the forward and backward chemical reaction rate constants for the $\ell^{th}$ reaction. 

Let $\mu_{j}^Y = \mu_{j}^{0,Y}+k_BT \log y_{j}$ be the chemical potential for $Y_{j}$, in which $\mu_{j}^{0,Y}$ is the standard chemical potential. Similar definitions also hold for $\mu_{i}^X$. Then for $1\leq\ell\leq M$,  one has \cite{Rao2016Nonequilibrium,SM}
\begin{equation} \label{eqstar_single}
    k_BT \log \frac{J^{+\ell}}{J^{-\ell}} = \Delta G_\ell=G_{+\ell}-G_{-\ell},
\end{equation}
in which 
$$G_{+\ell}=\sum_{i=1}^{N_{1}}\Xi_{+\ell}^{i, X}\mu_{i}^{X}+\sum_{j=1}^{N_{2}}\Xi_{+\ell}^{j, Y}\mu_{j}^{Y}$$
and
$$G_{-\ell}=\sum_{i=1}^{N_{1}}\Xi_{-\ell}^{i, X}\mu_{i}^{X}+\sum_{j=1}^{N_{2}}\Xi_{-\ell}^{j, Y}\mu_{j}^{Y}$$
are the total free energies of the reactants and products of the $\ell$-th reaction respectively. Therefore, Eq. \ref{eqstar_single} is exactly Eq. \ref{eqstar} for any single macroscopic chemical reaction.


On the other hand, the mesoscopic setting focuses on molecular numbers instead of molar concentrations in the CRN, which are stochastic. 
We use $\mathbf{n^Y}(t) = (n^Y_1(t),n^Y_2(t),\ldots,n^Y_{N_2}(t))$ to denote molecular numbers of the internal species at time $t$, which is a random vector.
X's species still possess fixed molecular numbers, denoted by $\mathbf{n^x} = (n^x_1,n^x_2,\ldots,n^x_{N_1})$.

Next it is necessary to describe how $\mathbf{n^Y}(t)$ evolves through time, which can be seen as a stochastic jumping process on a high-dimensional graph (called a \textit{counting space}). Each vertex of the graph has an $N_2-$dimensional coordinate $\mathbf{n^y}=(n^y_1, n^y_2,\ldots, n^y_{N_2})$ and if the occurrence of an elementary reaction in $\{R_1,...,R_M\}$ can convert the state of the system from one vertex to another, then there is an edge connecting them in this graph. 

The counting space is a scaffold for a Markov process with transition rates
$r_{+\ell}(\mathbf{n^y};V) \triangleq  \tilde{k}_{+\ell}(V) \prod_{j=1}^{N_2}n^y_{j}(n^y_{j}-1)\cdots(n^y_{j}-\Xi_{+\ell}^{j,Y}+1)$ for the forward reaction $R_{+\ell}$
($r_{-\ell}(\mathbf{n^y};V) \triangleq  \tilde{k}_{-\ell}(V) \prod_{j=1}^{N_2}n^y_{j}(n^y_{j}-1)\cdots(n^y_{j}-\Xi_{-\ell}^{j,Y}+1)$ for the corresponding backward reaction $R_{-\ell}$), except two reactions involving $X$, i.e. $r_{+1}(\mathbf{n^x},\mathbf{n^y};V) \triangleq  \tilde{k}_{+1} \prod_{i=1}^{N_1}n^x_{i}(n^x_{i}-1)\cdots(n^x_{i}-\Xi_{+1}^{i,X}+1) \prod_{j=1}^{N_2}n^y_{j}(n^y_{j}-1)\cdots(n^y_{j}-\Xi_{+1}^{j,Y}+1)$ and $r_{-M}(\mathbf{n^x}, \mathbf{n^y};V) \triangleq \tilde{k}_{-M} \prod_{i=1}^{N_1}n^x_{i}(n^x_{i}-1)\cdots(n^x_{i}-\Xi_{-M}^{i,X}+1)\prod_{j=1}^{N_2}n^y_{j}(n^y_{j}-1)\cdots(n^y_{j}-\Xi_{-M}^{j,Y}+1)$. Here, $\tilde{k}_{\pm \ell}$ are the mesoscopic \textit{rate constants} in the stochastic model, and $k_{\pm \ell} = \tilde{k}_{\pm \ell} V^{n_{\pm \ell}-1}$ where $n_{\pm \ell}$ is the summation of stoichiometric numbers of reactants in the reaction $R_{\pm \ell}$.


Then the chemical master equation(CME) describing the evolution of  the probability $p_V(\mathbf{n^y},t)=Prob(\mathbf{n^Y}(t) = \mathbf{n^y})$ is 
\begin{align*}
\frac{\partial p_V(\mathbf{n^y}, t)}{\partial t} &= \sum_{\ell=2}^{M-1} [r_{+\ell}(\mathbf{n^y} - \Xi_{\ell}^{Y};V)p_V(\mathbf{n^y} - \Xi_{\ell}^{Y}, t) \\
                                                 &- (r_{+\ell}(\mathbf{n^y};V)+r_{-\ell}(\mathbf{n^y};V))p_V(\mathbf{n^y}, t) \\
                                                 &+ r_{-\ell}(\mathbf{n^y} + \Xi_{\ell}^{Y};V)p_V(\mathbf{n^y} + \Xi_{\ell}^{Y}, t)]\\
                                                 &+ [r_{+1}(\mathbf{n^x},\mathbf{n^y} - \Xi_{1}^{Y};V)p_V(\mathbf{n^y} - \Xi_{1}^{Y}, t) \\
                                                 &- (r_{+1}(\mathbf{n^x},\mathbf{n^y};V)+r_{-1}(\mathbf{n^y};V))p_V(\mathbf{n^y}, t) \\
                                                 &+ r_{-1}(\mathbf{n^y} + \Xi_{1}^{Y};V)p_V(\mathbf{n^y} + \Xi_{1}^{Y}, t)]\\
                                                  &+ [r_{+M}(\mathbf{n^y} - \Xi_{M}^{Y};V)p_V(\mathbf{n^y} - \Xi_{M}^{Y}, t) \\
                                                 &- (r_{+M}(\mathbf{n^y};V)+r_{-M}(\mathbf{n^x},\mathbf{n^y};V))p_V(\mathbf{n^y}, t) \\
                                                 &+ r_{-M}(\mathbf{n^x},\mathbf{n^y} + \Xi_{M}^{Y};V)p_V(\mathbf{n^y} + \Xi_{M}^{Y}, t)],
\end{align*}
where $\Xi_{\ell}^{Y}$ is just the $\ell$-th column of the matrix $\bold{\Xi}^Y$ with $1\leq \ell \leq M$.



{\em Kinetic cycle, net cycle flux and one-way cycle fluxes.} A directed cycle in the counting space is a slice of path with the same origin and destination in which no other states overlap. It can be expressed in this way: $c= [\mathbf{y}_1,\mathbf{y}_2, \ldots, \mathbf{y}_k]$ with some integer $k$ where $\mathbf{y}_1,\mathbf{y}_2,\ldots, \mathbf{y}_k$ are successive states in the path which are different from each other. The next state of $\mathbf{y}_k$ in the path is exactly $\mathbf{y}_1$, forming a single closed loop. We identify $[\mathbf{y}_1,\mathbf{y}_2,\ldots, \mathbf{y}_k]$ and $[\mathbf{y}_i,\mathbf{y}_{i+1},\ldots, \mathbf{y}_{i+k-1}]$ for any $i$ with indices modulo $k$, as the same directed cycle. 

According to the theory of cycle representation and cycle fluxes for Markov process \citep{Jiang20041, Jiang20042, Kalpazidou1995Cycle}, one can trace along every sample path, and calculate the time-averaged number of times that a particular cycle $c$ formed by time $t$. Then letting the time $t$ go to infinity the limit defines the cycle flux $\omega_c$ for $c$.
In the CME, the system is at equilibrium (time-reversible) if and only if $\omega_c = \omega_{c_-}$ for every cycle $c$ \cite{Jiang20041,Jiang20042}, in which $c_-$ is the reversed cycle of $c$. Furthermore, the ratio 
\begin{equation}\label{ratio_conf_cycle}
\frac{\omega_c}{\omega_{c_-}}=\frac{r_{\ell_1}r_{\ell_2}\cdots r_{\ell_k}}{r_{-\ell_1}r_{-\ell_2}\cdots r_{-\ell_k}}
\end{equation}
where $\{\ell_1,\ell_2,\ldots,\ell_k\}$ represents the successive reactions $\{R_{\ell_1},R_{\ell_2},\ldots,R_{\ell_k}\}$ occurred in the cycle $c= [\mathbf{y}_1,\mathbf{y}_2,\ldots, \mathbf{y}_k]$, with $-\ell_i$ being the reversed reaction of the reaction $\ell_i$ and $r_{\ell}$ is the corresponding transition rates. The exact expression of cycle fluxes and the corresponding cycle decomposition are given in \cite{SM}, cited from \cite{Jiang20041,Jiang20042}.



However, in many applications one is interested in the flux along reaction cycles rather than the cycle fluxes for any given cycle in the counting space.
More specifically, such cycles, which we call \textit{reaction cycles}, are right null vectors for the matrix $\Xi^Y$ \cite{polettini2014irreversible}. We denote such a cycle as $\tilde{c} = (c_1,c_2,\ldots,c_M)$ in which $c_\ell$ is the net number of the occurrence for the reaction $R_\ell$. All the reaction cycles form a vector space. 

Furthermore we can prove that the ratio $c_1/c_M$ is a positive invariant number \cite{SM}, the simplest form of which is denoted as $\frac{c_{in}}{c_{out}}$. Therefore, we assume $c_1 = n c_{in}$ and $c_M = n c_{out}$, where $n \in \mathbb{Z}$. The overall effect of a reaction cycle is just to make 
$\sum_{i} c_{in} \Xi_{+1}^{i,X} X_{i}\rightarrow\sum_{i}c_{out} \Xi_{-M}^{i,X} X_{i}$, which is just from the left-hand-side of reaction $R_1$ to the right-hand-side of reaction $R_M$ in Fig. \ref{fig:1}, occurring $n$ times. Once $n=0$, then both $c_1=c_M=0$, it is a \textit{closed cycle}.


For each cycle $c$ in the counting space, the net numbers of occurrence of all the reactions in $c$ form a reaction cycle $\tilde{c}$. It follows naturally that the cycle flux of $\tilde{c}$ is defined as the sum of the cycle fluxes of all the counting cycles $c$ assorted to it. More formally we can define a map $\phi: \mathcal{C}_{\infty} \rightarrow \mathbb{R}^M$ with 
$\mathcal{C}_{\infty}$ being the set of all cycles in the counting space and $\phi(c)$ being the unique reaction cycle generated by $c$ as above.  Then the cycle flux for $\tilde{c}$ can be interpreted as the averaged frequency $\tilde{c}$ occurs in the CME model. More precisely, the reaction cycle flux of $\tilde{c}$
$$\omega_{\tilde{c}} = \sum_{\phi(c) = \tilde{c}} \omega_c.$$

It's also easy to verify that if a cycle $c$ is assigned to $\tilde{c}$, then its time reversal $c_-$ will be assigned to 
$-\tilde{c}$. We thus just define the reversed reaction cycle $\tilde{c}_-$ of $\tilde{c}$ as $-\tilde{c}$. Notice that for each cycle assigned to the same reaction  cycle, the ratio of its flux and the flux of the corresponding reversed cycle is only dependent on how many times each reaction occurred in the cycle regardless of the order, being the same \cite{SM}. Therefore, we have \cite{SM}
\begin{align}\label{1stresult}
\frac{\omega_{\tilde{c}}}{\omega_{\tilde{c}_-}}& =\prod_{\ell = 1}^{M} \frac{\tilde{k}_{+\ell}^{c_\ell}}{\tilde{k}_{-\ell}^{c_\ell}} \prod_{i = 1}^{N_1} \frac{(n^x_{i}(n^x_{i}-1)\cdots(n^x_{i}-\Xi_{+1}^{\i,X}+1))^{c_1}}{(n^x_{i}(n^x_{i}-1)\cdots(n^x_{i}-\Xi_{-M}^{i,X}+1))^{c_M}}.
\end{align}

Once $c_1=c_M=0$, i.e. closed cycle, the ratio is 1 according to the Wegscheidier's Condition, i.e. $\prod_{\ell = 2}^{M-1} \frac{\tilde{k}_{+\ell}^{c_\ell}}{\tilde{k}_{-\ell}^{c_\ell}} = 1$, indicating no driving force available for the cycle. In a closed system with only closed cycles, where there are no input and output reactions $R_1$ and $R_M$ present, $\frac{\omega_{\tilde{c}}}{\omega_{\tilde{c}_-}} = 1$ for each reaction  cycle, which is the hallmark of equilibrium state.

Eq. \ref{1stresult} is the first main result of this letter, which leads to the generalization of Eqs. \ref{eqstar} and \ref{eqstar_single} for each reaction cycle (See Eq. \ref{macro-force-flux1}). Typically, the external species $X$ is controlled macroscopically in the unit of molar concentration $x_{i} = V^{-1}n_{i}^x$. Then since $k_{\pm \ell} = \tilde{k}_{\pm \ell} V^{n_{\pm \ell}-1}$, and defining the macroscopic cycle currents for the reaction cycle $\tilde{c}$ as $\mathcal{J}_{\tilde{c}} = \lim_{V \to \infty}\frac{\omega_{\tilde{c}}}{V}$, when $V \to \infty$, Eq. \ref{1stresult} becomes \cite{SM}
\begin{equation}
\frac{\mathcal{J}_{\tilde{c}}}{\mathcal{J}_{\tilde{c}_-}}=\left[\prod_{\ell = 1}^{M} \frac{k_{+\ell}^{c_\ell}}{k_{-\ell}^{c_\ell}}\right] 
\cdot \left[\prod_{i=1}^{N_1} (x_{i})^{c_1\Xi_{+1}^{i,X} - c_M\Xi_{-M}^{i,X}}\right].
\end{equation}

We can also obtain the corresponding cycle decomposition of the macroscopic fluxes $J^{\ell}$ in terms of the reaction cycles (more detailed decomposition is derived in \cite{SM}):
\begin{equation}\label{cycle_decomposition_macro}
    J^{\ell}=\sum_{\tilde{c} \in \phi(\mathcal{C}_{\infty})} c_{\ell}\cdot\mathcal{J}_{\tilde{c}}.
\end{equation}

This is exactly the relation that used for the definition of reaction-cycle currents in \cite{polettini2014irreversible,Rao2016Nonequilibrium}, which is not unique. Here in our definition, all the reaction-cycle fluxes are exactly their averaged frequency of occurrence per unit time at steady state, and we are able to distinguish the cycle fluxes of the original cycles and their reversed ones, making the definition of one-way fluxes plausible\cite{SM}. 

Notice that $k_BT\log\prod_{\ell = 1}^{M} \frac{k_{+\ell}^{c_\ell}}{k_{-\ell}^{c_\ell}}$ is equal to $n\Delta G^0_{1\rightarrow M}$, in which $\Delta G^0_{1\rightarrow M}=\sum_{i} \left(c_{in}\Xi_{+1}^{i,X}-c_{out} \Xi_{-M}^{i,X}\right) \mu^{0,X}_{i}$ is the intrinsic free energy difference of the overall reaction $ \sum_{i} c_{in}\Xi_{+1}^{i,X} X_{i}\rightarrow\sum_{i} c_{out}\Xi_{-M}^{i,X} X_{i}$ \cite{SM}, then
\begin{equation}\label{macro-force-flux1}
    k_BT\log\frac{\mathcal{J}_{\tilde{c}}}{\mathcal{J}_{\tilde{c}_-}}= n\Delta G_{1\rightarrow M},
\end{equation}
in which $\Delta G_{1\rightarrow M}=\sum_{i} \left(c_{in}\Xi_{+1}^{i,X}-c_{out} \Xi_{-M}^{i,X}\right) \mu^X_{i}$ is exactly the free energy difference of the overall reaction and $\mu^X_{i}$ is the chemical potential of $X_{i}$ \cite{SM}.

Once a reaction cycle is completed, a certain number ($n$) of $\sum_{i} c_{in}\Xi_{+1}^{i,X} X_{i}$ is transformed into $\sum_{i} c_{out}\Xi_{-M}^{i,X} X_{i}$, which is the overall effect of a reaction cycle. Many reaction cycles share the same overall effect. So we can define the one-way flux upon the overall effect, namely the value of $n$ in this single input/output scenario, i.e.
\begin{equation}
J_n = \sum_{c_1=nc_{in}} \omega_{\tilde{c}},~\forall n \in \mathbb{Z}.\nonumber
\end{equation}



Actually, $\log\prod_{\ell = 1}^{M} \frac{\tilde{k}_{+\ell}^{c_\ell}}{\tilde{k}_{-\ell}^{c_\ell}}$ in Eq. \ref{1stresult} is also a linear function only depends on the value of $n$, being the same for all the reaction cycles with the same overall effect \cite{SM}, denoted as $n\cdot\kappa$. Therefore
\begin{align*}
\frac{J_n}{J_{-n}} &=\frac{\sum_{c_1=n c_{in}}\omega_{\tilde{c}}}{\sum_{c_1=-n c_{in}}\omega_{\tilde{c}}}\\
				&= e^{n\kappa} \prod_{i = 1}^{N_1} \frac{(n^x_{i}(n^x_{i}-1)\cdots(n^x_{i}-\Xi_{+1}^{i,X}+1))^{n c_{in}}}{(n^x_{i}(n^x_{i}-1)\cdots(n^x_{i}-\Xi_{-M}^{i,X}+1))^{n c_{out}}}.
\end{align*}


Take $V \to \infty$, we also obtain
$$k_BT\log\frac{\mathcal{J}_n}{\mathcal{J}_{-n}}= n\Delta G_{1\rightarrow M},$$
where we use the similar notation $\mathcal{J}_n = \lim_{V \to \infty} \frac{J_n}{V}$ for the macroscopic one-way flux dependent only on the overall effect.


{\em Steady-state entropy production rate} The steady-state mesoscopic entropy production rate (mesoEPR) in the counting space can be expressed by all the cycle fluxes \cite{Jiang20042}
$$mesoEPR=\frac{1}{2}k_BT\sum_{c}(\omega_c-\omega_{c_{-}})\log\frac{\omega_c}{\omega_{c_{-}}}.$$

In terms of reaction cycles, the reaction-cycle decomposition of $mesoEPR$ is straightforwardly obtained:
\begin{equation}\label{MesoRecCyc}
    mesoEPR= \frac{1}{2} k_BT \sum_{\tilde{c}} (\omega_{\tilde{c}} - \omega_{\tilde{c}_-}) \log \frac{\omega_{\tilde{c}}}{\omega_{\tilde{c}_-}}=\sum_{\tilde{c}} \omega_{\tilde{c}} \mathcal{A}^{meso}_{\tilde{c}},
\end{equation}
in which $\mathcal{A}^{meso}_{\tilde{c}}=k_BT \log \frac{\omega_{\tilde{c}}}{\omega_{\tilde{c}_-}}$.

By the way, the $mesoEPR$ can also be decomposed by the overall effect of reaction cycles:
\begin{align}\label{MesoOneway}
mesoEPR&=  \frac{1}{2}k_BT \sum_{n=-\infty}^{\infty}\sum_{c_1=nc_{in}} (\omega_{\tilde{c}} - \omega_{\tilde{c}_-}) \log \frac{\omega_{\tilde{c}}}{\omega_{\tilde{c}_-}}\nonumber\\
&= \frac{1}{2}k_BT \sum_{n=-\infty}^{\infty}(J_n-J_{-n}) \log \frac{J_n}{J_{-n}}\nonumber\\
&= \frac{1}{2}k_BT \sum_{n=-\infty}^{\infty} n(J_n-J_{-n})\log \frac{J_1}{J_{-1}}\nonumber\\
&= k_BT J_T \log \frac{J_1}{J_{-1}},
\end{align}
noticing $\frac{J_n}{J_{-n}}=\left[\frac{J_1}{J_{-1}}\right]^n$ and $J_T = \sum_{n=-\infty}^{\infty}nJ_n$ is the total net flux of $X$.

Towards the macroscopic scale as the volume goes to infinity, it has been recently proved that \cite{Ge2017Mathematical}, the limit of $mesoEPR/V$ is exactly the steady-state macroscopic entropy production rate
$$macroEPR= \frac{1}{2}k_BT\sum_{\ell=-M, \ell\neq 0}^M(J^{\ell}-J^{-\ell})\log\frac{J^{\ell}}{J^{-\ell}}.$$

The cycle decomposition of $macroEPR$ can be derived from Eq. \ref{cycle_decomposition_macro} \cite{Rao2016Nonequilibrium,polettini2014irreversible,SM}, i.e. 
\begin{equation}
macroEPR=\sum_{\tilde{c}}\mathcal{J}_{\tilde{c}}\mathcal{A}^{macro}_{\tilde{c}}=\mathcal{J}_T\Delta G_{1\rightarrow M},
\end{equation}
in which $\mathcal{A}^{macro}_{\tilde{c}} = k_BT \log \frac{\mathcal{J}_{\tilde{c}}}{\mathcal{J}_{\tilde{c}_-}} = k_BT\sum_{\ell=1}^Mc_{\ell}\ln\frac{J^{+\ell}}{J^{-\ell}}$ is just the the change of Gibbs free energy along $\tilde{c}$, called cycle affinity, and $\mathcal{J}_T = \sum_{n=-\infty}^{\infty}n\mathcal{J}_n$ is the total net macroscopic flux. It is easy to see that the limit of $\mathcal{A}^{meso}_{\tilde{c}}$ is exactly $\mathcal{A}^{macro}_{\tilde{c}}$. So we can also obtain the decomposition of $macroEPR$ by limiting $mesoEPR/V$ in (\ref{MesoRecCyc}) \cite{SM}.

{\em Generalization to general CRN with multiple input and output complexes} Generally the stoichiometric matrix shall have the structure
\begin{equation}
\Xi=\begin{bmatrix}
\Xi_b^X & 0 \\
\Xi_b^Y & \Xi_i^Y
\end{bmatrix},    
\end{equation}
in which $\Xi_b^X$  and $\Xi_b^Y$ indicate the coefficients of reactions ``on the boundary'', namely involving the chemostatted $X$'s, and $\Xi_i^Y$ is for the reactions among internal species $Y$'s. 

Eq. \ref{ratio_conf_cycle} also holds for each cycle in the counting space. Then for each reaction cycle $\tilde{c}$, which is still the right null vector of $[\Xi_b^Y, \Xi_i^Y]$, also define $\omega_{\tilde{c}} = \sum_{\phi(c) = \tilde{c}} \omega_c$, then the cycle affinity
\begin{equation}
    \mathcal{A}^{macro}_{\tilde{c}}=\lim_{V\to \infty}\mathcal{A}^{meso}_{\tilde{c}}= k_BT\log\frac{\mathcal{J}_{\tilde{c}}}{\mathcal{J}_{\tilde{c}_-}}=\Delta G_{\tilde{c}}=-\vec{l}\cdot\mu_X,
\end{equation}
in which $\vec{l}=[\Xi_b^X,0]\times\tilde{c}$ is the overall effect of $\tilde{c}$ and $\mu_X=(\mu^X_{i})$ is the chemical potential of external species $X$. $\Delta G_{\tilde{c}}$ is the free energy difference along the reaction cycle $\tilde{c}$, which is only dependent on the overall effect $\vec{l}$. It is the force-flux relation under the most general setting to date.

Therefore, we can also group all the reaction cycles with the same overall effect together, defining the mesoscopic one-way flux $J_{\vec{l}}$ and the macroscopic one-way flux $\mathcal{J}_{\vec{l}}$. The ratio of the macroscopic one-way flux $J_{\vec{l}}$ with overall effect $\vec{l}$ and $J_{-\vec{l}}$ with  $-\vec{l}$ satisfies
\begin{equation}\label{ForceFluxGen}
    k_BT\log\frac{J_{\vec{l}}}{J_{-\vec{l}}}=-\vec{l}\cdot\mu_X\stackrel{\triangle}{=}\Delta G_{\vec{l}}.
\end{equation}


The steady-state mesoscopic entropy production rate (mesoEPR) in the counting space keeps several levels of cycle decomposition, i.e.
\begin{eqnarray}
&&mesoEPR = \frac{1}{2}k_BT\sum_{c}(\omega_c-\omega_{c_{-}})\log\frac{\omega_c}{\omega_{c_{-}}}
\nonumber\\
&=& \frac{1}{2} k_BT \sum_{\tilde{c}} (\omega_{\tilde{c}} - \omega_{\tilde{c}_-}) \log \frac{\omega_{\tilde{c}}}{\omega_{\tilde{c}_-}}= \sum_{\tilde{c}} \omega_{\tilde{c}} \mathcal{A}^{meso}_{\tilde{c}}
\nonumber\\
&=& \frac{1}{2}k_BT \sum_{\vec{l}}(J_{\vec{l}}-J_{-\vec{l}}) 
     \log \frac{J_{\vec{l}}}{J_{-\vec{l}}}= \sum_{\vec{l}}J_{\vec{l}}\Delta G_{\vec{l}}
\end{eqnarray}

Taking $V$ goes to infinity, 
the steady-state macroscopic entropy production rate
\begin{equation}
    macroEPR =\sum_{\tilde{c}}\mathcal{J}_{\tilde{c}}\mathcal{A}^{macro}_{\tilde{c}}.
\end{equation}

If we further define the macroscopic one-way flux for overall effect $\vec{l}$ as $\mathcal{J}_{\vec{l}} = \lim_{V \to \infty} \frac{J_{\vec{l}}}{V}$ and using (\ref{ForceFluxGen}), we can obtain 
\begin{align*}
   macroEPR =\sum_{\vec{l}}\mathcal{J}_{\vec{l}}\Delta G_{\vec{l}} = - (\sum_{\vec{l}}\mathcal{J}_{\vec{l}} \;  \vec{l}) \cdot \mu_X = - j_X \cdot \mu_X
\end{align*}
where $j_X$ denotes the overall steady-state flux of $X$'s caused by the system. This is an acknowledged formula in metabolic network \cite{Bonarius1997Flux,Orth2010What,Beard2002Energy}.

All the details of our derivation can be found in the SI \cite{SM}.

{\em Conclusion and discussion} Using the theory of cycle representation for Markov processes, mesoscopic and macroscopic one-way cycle fluxes
in a nonlinear complex CRN are rigorously defined. Cycle decomposition are carried out at three different levels, including mesoscopic counting space, emergent cycle space, and macroscopic CRN; the force-flux relation is demonstrated for each reaction cycle.  These provided a consistent computation for entropy production. The universal relation between one-way fluxes and thermodynamic driving force proven in the present work underpins a consistency between kinetics and thermodynamics. It introduces new perspectives into and builds bridges between theories of CRNs at different scales. 


\bibliographystyle{unsrt}
\bibliography{project}
\end{document}